\documentclass{article}
\usepackage{RR}
\usepackage[latin9]{inputenc}
\usepackage{graphicx,xspace}
\usepackage[T1]{fontenc} 
\usepackage{parskip} \parskip=1ex plus 5pt minus 1pt \parindent=4ex 
\usepackage{amsfonts,amsmath} 
\usepackage{amssymb,amsthm}
\usepackage{eurosym} 
\usepackage[obeyspaces]{url} 
\usepackage{hyperref}  

\usepackage{local}

\RRdate{9 May 2008}
\RRNo{6546}

\newcommand{\daemon}{d{\ae}mon\xspace}

\RRtitle{Telex\,: un système de partage en écriture pour les
  applications collaboratives, basé sur un modèle formel}
\RRetitle{{Telex: Principled System Support for Write-Sharing in
    Collaborative Applications}%
\thanks{
    This research is supported in part by Respire (ANR, France,
    \url{respire.lip6.fr}), Grid4All (FP6, EU,
    \url{www.grid4all.eu}) and grant JC2007-00213 (Spain).
}
}
\titlehead{Telex: Principled Sys.\ Support for Write-Sharing in Collab.\ Apps.}

\RRauthor{
  Lamia Benmouffok\thanks[inria]{INRIA, Paris-Rocquencourt, France}%
                  \thanks[lip6]{LIP6, Paris, France}
\and                 
  Jean-Michel Busca\thanksref{inria}%
                   \thanksref{lip6}
\and
  Joan Manuel Marquès\thanks{Universitat Oberta de Catalunya, Barcelona, Spain}%
                     \thanksref{lip6}
\and
  Marc Shapiro\thanksref{inria}%
              \thanksref{lip6}
\and
  Pierre Sutra\thanksref{inria}%
              \thanksref{lip6}
\and
  Georgios Tsoukalas\thanks{National Technical University of Athens, Greece}
}
\authorhead {Benmouffok \emph{et al.}}

\abstract{  
  The Telex system is designed for sharing mutable data in a 
  distributed environment, particularly for collaborative applications.
  Users operate on their local, persistent replica of shared documents;
  they can work disconnected and suffer no network latency.
  The Telex approach to detect and correct conflicts is application
  independent, based on an action-constraint graph (ACG) that summarises
  the concurrency semantics of applications.
  The ACG is stored efficiently in a \emph{multilog} structure that
  eliminates contention and is optimised for locality.
  Telex supports multiple applications and multi-document updates.
  The Telex system clearly separates system logic (which includes
  replication, views, undo, security, consistency, conflicts, and
  commitment) from application logic.
  An example application is a shared calendar for managing multi-user
  meetings; the system detects meeting conflicts and resolves them
  consistently.
}
\resume{
  Le système Telex est conçu pour le partage des données modifiables
  dans un environnement réparti, principalement pour des applications
  collaboratives.
  Les utilisateurs opèrent sur une copie locale et persistante des
  documents qu'ils partagent\; ils peuvent travailler en mode
  déconnecté, et ne sont pas ralentis par la latence du réseau.
  Telex utilise une approche indépendante de l'application pour détecter
  et corriger les conflits, qui se base sur un graphe
  actions-contraintes (ACG) qui résume la sémantique de concurrence des
  applications.
  L'ACG est stocké de façon efficace dans une structure dite
  \emph{multi-journal} qui élimine la contention et est optimisée pour
  la localité.
  Des applications différentes s'exécutent sur Telex, qui permet de
  mettre à jour plusieurs documents de façon coordonnée.
  Telex sépare proprement la logique système (ce qui inclut la
  réplication, les vues, le «undo», la sécurité, la cohérence, les
  conflits, et la finalisation) de la logique applicative.
  Un exemple d'application est un calendrier partagé, pour gérer des
  réunions multi-utilisateur\; le système détecte les conflits de
  réunion et les résout de façon cohérente.
}

\RRequipe{Regal}
\RRtheme{\THCom}
\RCParis

\begin{document}
\makeRR

\sloppy

\section{Introduction}
\label{sec:introduction}

The Telex system provides novel solutions for write-sharing data in
co-operative and disconnected work settings.



Existing approaches have severe limitations.
For instance state machine replication \cite{con:rep:615} imposes high
latency and does not support disconnected operation.
The popular last-writer-wins algorithm \cite{optim:rep:syn:1500} does
not ensure any high-level correctness guarantees.%
\footnote{
    Section~\ref{sec:related} analyses the state
    of the art in detail.
}
In contrast, Telex is based on a principled approach that combines
flexibility and correctness, and cleanly separates application logic
from system logic.


Application logic transmits to Telex actions (operations) and
constraints (concurrency invariants), and applies execution schedules
transmitted by Telex.
In return, Telex takes care of: replication, consistency, storage and
access control; collecting, transmitting and persisting operations; 
detecting conflicts and
computing high-quality conflict-free schedules; forward execution and
rollback; checkpointing; commitment; and access control.
Telex supports multi-document updates and cross-application scenarios
out of the box.

Telex is based on a principled approach, the Action-Constraint Graph
(ACG) \cite{optim:rep:syn:1498}.
We designed the \emph{multilog} data structure to store ACG-based
documents in a distributed file system.
Multilogs eliminate write contention and promote locality.


We developed a number of demonstration applications above Telex.
For instance, a shared calendar application lets people organise their
agenda collaboratively, arranging private events and group meetings.
Telex detects meeting conflicts and proposes possible solutions.


The contributions of this paper include:
a novel approach to shared data replication that is application
independent yet application-aware, the ACG;
the practical engineering of an ACG system, in particular the
document and multilog structures;
design examples and lessons learned for ACG-based applications;
and
some benchmarks and performance measurements.

This paper proceeds as follows.
Section~\ref{sec:overview} is an overview.
Section~\ref{sec:structures} explains the data structures that Telex uses.
Section~\ref{sec:architecture} documents the Telex architecture and
implementation.
In Section~\ref{sec:applications}, we present some example applications.
Section~\ref{sec:evaluation} evaluates the Telex performance.
We reflect on lessons learned in Section~\ref{sec:lessons}.
Section~\ref{sec:related} compares Telex with related work.
Finally, Section~\ref{sec:conclusion} concludes.

\begin{figure*}[t]
  \centering\footnotesize
  \begin{tabular}{@{}ccc@{}}
    \includegraphics[width=4cm]{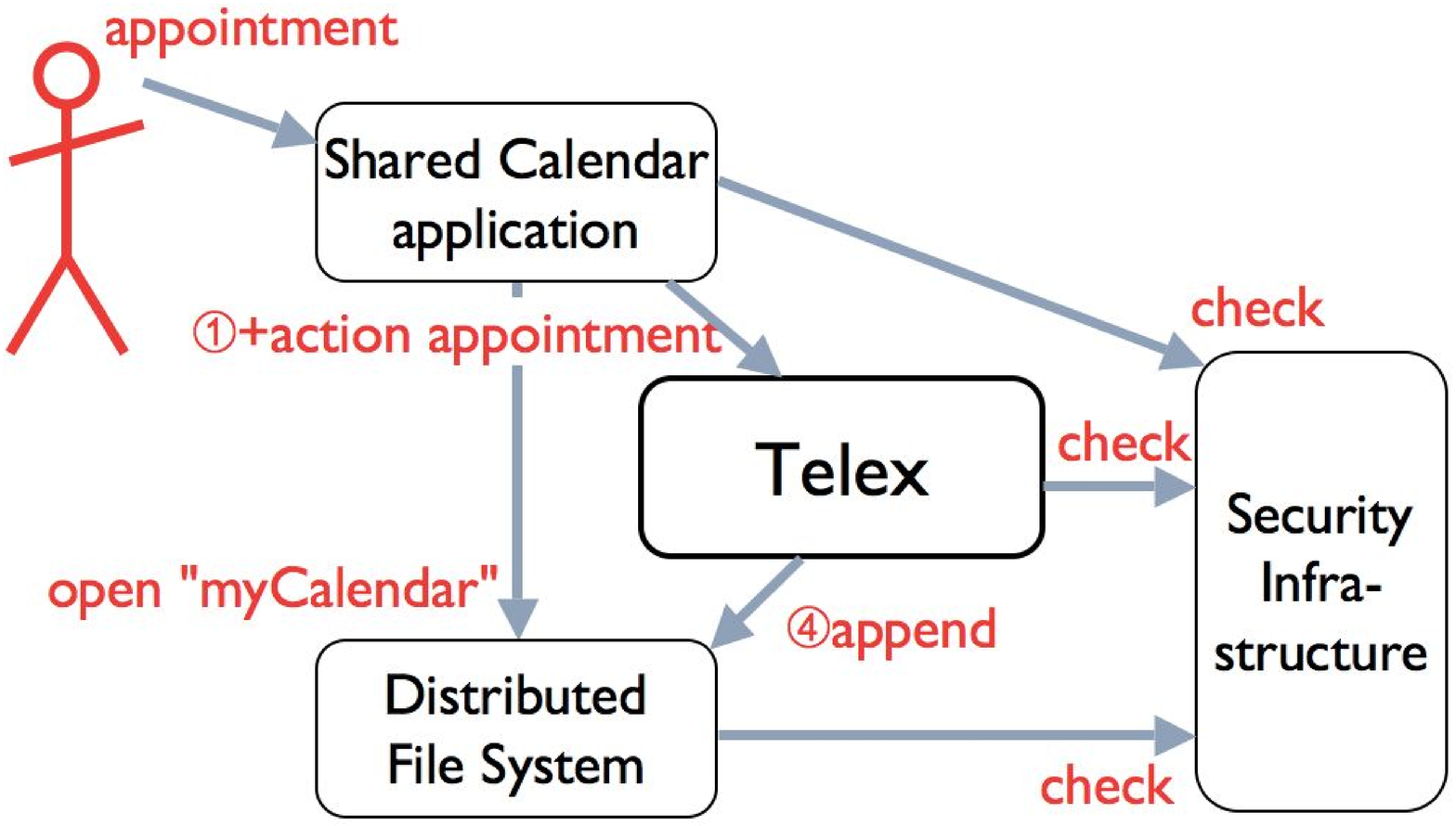}
    &
    \includegraphics[width=4cm]{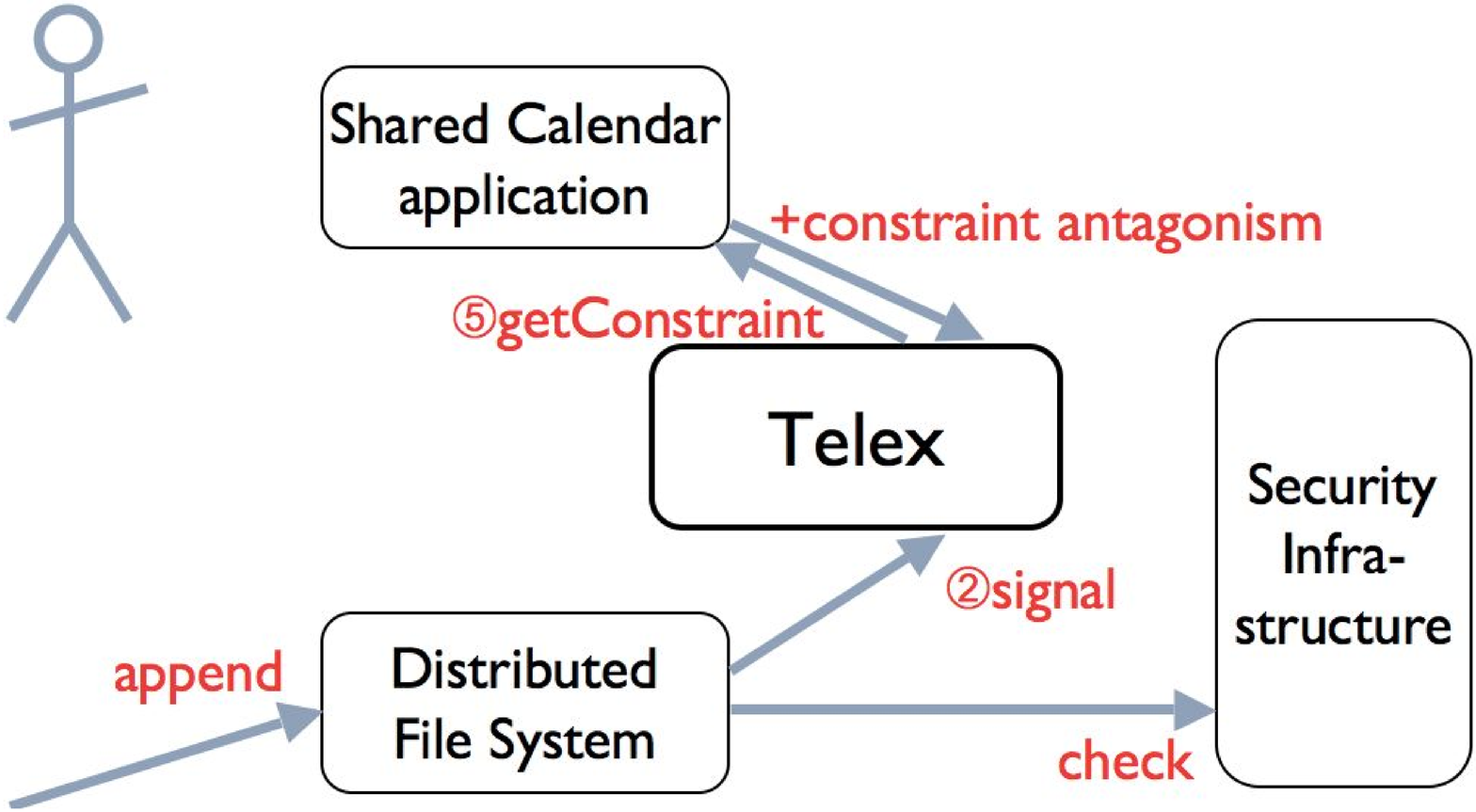}
    &
    \includegraphics[width=4cm]{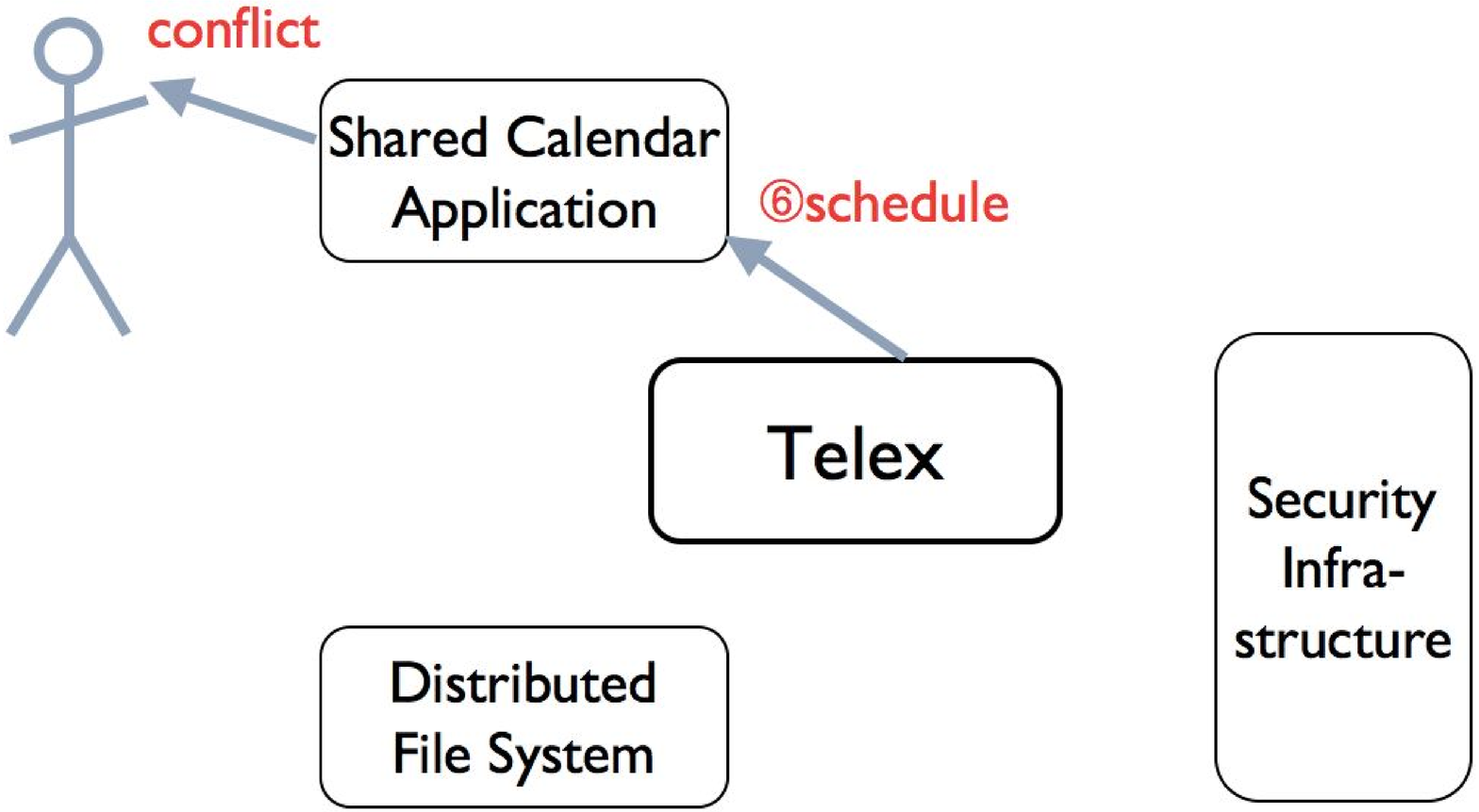}
\\ & & \\
    {a. App.\ reifies user op.}
&
    {b. Remote action rcvd.:}
&
    {c. Compute schedule(s),}
\\
    {as actions \& constraints}
&
    {upcall for conflict constraints}
&
    {execute, display}
\\
  \end{tabular}
  \caption{Telex interactions.  (The circled numbers refer to
    Figure~\ref{fig:telexArchitecture})}
  \label{fig:app-scenario}
\end{figure*}

\section{Telex overview}
\label{sec:overview}

We give an overview of the Telex system from three complementary points
of view.

\subsection{User{\slash}application perspective}


Telex supports \emph{participants}, i.e., users working at disjoint
\emph{sites}, which may be widely distributed.
An authorised participant may replicate a shared \emph{document} on his
site.

A site operates optimistically \cite{optim:rep:syn:1500}: it applies
local \emph{actions} (operations), sends them to other sites, and
eventually \emph{replays} the actions it receives.
Hence, applications are not slowed down by remote synchronisation,
network issues, or by remote failures.

A participant may work either connected or disconnected from others.
Thus, each participant has his own \emph{view} of the current state of the
shared document.
Documents and views persist across log-out{\slash}log-in and
restarts.
However, a view is only tentative and may have to roll back.


Telex, not applications, takes care of hard issues such as conflict
detection, reconciliation, and consistency.
However, since a conflict is the violation of some application
invariant, Telex is parameterised by application-specific concurrency
invariants called \emph{constraints}.
A constraint relates two actions, either of the same or distinct
documents.
Hence, Telex maintains consistency between documents.


Figure~\ref{fig:app-scenario} illustrates the control structure of
Telex with a Shared Calendar (SC) application.%
\footnote{
    Elements of the figure not discussed here will be explained in later
    sections.
}
In this example, the participant creates an appointment, which conflicts
(double booking) with one
created remotely.
In Figure~\ref{fig:app-scenario}.a, the participant performs the
\textsf{appointment} operation.
The SC application logs the corresponding actions and constraints to the
local Telex d\ae{}mon (\textsf{+action appointment}).
In Figure~\ref{fig:app-scenario}.b, when the site receives a remote
action (\textsf{signal}), it compares it to the concurrent actions.
If Telex suspects a conflict, it calls up to the application
(\textsf{getConstraint}), which replies with precise information
(\textsf{+constraint antagonism}).
Finally, as in Figure~\ref{fig:app-scenario}.c, Telex periodically sends
\emph{schedules} to the application, for execution and{\slash}or
rollback.
The application computes and displays the corresponding views, in this
example with a conflict indication (\textsf{conflict}).%
\footnote{
    For the purpose of this paper, document state, view and schedule are
    synonymous.
    ``View'' emphasises that the state is local and is not unique;
    ``schedule'' emphasises that it is computed by some ordering of
    available actions.
}

\begin{table}[t]
  \centering\scriptsize
  \begin{tabular}{@{}c|c|c@{}}
    Name             & Notation            & Semantics                                                \\
\hline
    \NotAfterTxt     & $A \NotAfter B$     & $A$ is never after $B$ in any schedule                   \\
    \EnablesTxt      & $A \Enables B$      & $B$ in a schedule implies $A$ in same schedule           \\
    \NonCommutingTxt & $A \NonCommuting B$ & Must agree on $A \NotAfter B$ or $B \NotAfter A$ (conflict)  \\
\hline
    \AtomicTxt       & $A \Atomic B$       & All or nothing                                           \\
    \CausalTxt       & $A \Causal B$       & $B$ depends causally on $A$                              \\
    \AntagonismTxt   & $A \Antagonism B$   & $A$ and $B$ never both in same schedule (conflict) \\
  \end{tabular}
  \caption{Constraints}
  \label{tab:constraints}
\end{table}

\subsection{Formal perspective: actions and constraints}
\label{sec:formal}


Telex is based on a formal model, the Action-Constraint Graph (ACG)
\cite{optim:rep:syn:1498}.
The ACG is a labelled graph whose nodes are the actions and edges are the
constraints.
The current view of a site is the result of executing a \emph{sound
  schedule}, i.e., an ordering of actions currently known at that site,
that obeys the safety constraints \NotAfterTxt and \EnablesTxt.
In effect, the ACG represents the set of all legal views.

Table~\ref{tab:constraints} presents briefly the constraints supported
by Telex; for full details please refer to the
relevant publications \cite{optim:rep:syn:1498}.
The first three are primitive, the last three are combinations of the
primitives.%
\footnote{
    \AtomicTxt does not ensure transactional isolation; an isolation
    constraint will be added in the future.
    Currently, to achieve isolation, the user must manually group
    operations into a single action.
}


These represent important classes of concurrency invariants.
While they can approximate the true application semantics only grossly,
we have found that they are sufficiently expressive for reconciliation
purposes in several kinds of applications
\cite{optim:syn:1474,rep:syn:1482}.


Formally, eventual consistency requires that all schedules be sound, that
they have a common stable sound prefix, that every action eventually be
either aborted or in the prefix, and that non-commuting actions that are
in the prefix be ordered.%
\footnote{
    Mutually-commuting actions may run in any relative order.
}
The latter two items imply a global consensus between sites.
We call this consensus the \emph{commitment protocol}.
In Telex, commitment is optimistic, i.e., it occurs in the background,
not in the critical path of applications.

\subsection{Engineering perspective: multi-logs and commitment}
\label{sec:engineering}

The design of Telex is motivated by some major requirements and challenges:
(i)    Persist and replicate the ACG.
(ii)   Provide strong guarantees above a distributed file system with only
       best-effort consistency.
(iii)  Integrate documents into the file system, with
       reasonable overhead and scalability.
(iv)   Provide access control, without violating consistency.
(v)    Remove old ACG entries from storage.
(vi)   Decentralised, peer-to-peer design,
       with support for casual disconnected operation.


A document is a named entity in the file system.
For locality, a document stores only the portion of the ACG consisting of the
actions operating on the document, and their constraints.

Telex documents coexist with ordinary files and directories in the
file system.
Using one or the other is up to the application.


Telex relies on external mechanisms to store and replicate documents, and to
propagate changes to remote sites.
To avoid file system bottlenecks and consistency issues, each
participant writes to a distinct append-only \emph{log} within a
document.
To enable incremental garbage collection, the log is broken down into
successive chunk files.
This structure is called \emph{multilog}.

A log is a succession of actions and constraints in no particular
order.
We optimise for the expected common case, where constraints are inside the
same log; inter-log constraints within the same document are slightly
more expensive.
Inter-document constraints are assumed to be relatively rare and are
more costly.





Because of network delays and disconnections, and because of 
filtering and access control (explained later), at any point in time,
different participants may observe different ACGs.
However, each participant's view is consistent, because it results from
a sound schedule.
Thus, if some action $A$ is not in a view, and $A~\EnablesTxt~B$, then
$B$ is also not in that view.

The current view can be recorded in a \emph{snapshot}.
Snapshots name a view, speed up the computation of later views, and help
with garbage collection.


A decentralised, background commitment protocol ensures that the common
prefix of schedules makes progress.
Each participant can vote for a schedule according, for instance, to
user preference.
Voting is decentralised and peer-to-peer.

Committed log records may be deleted.
However it may be advantageous to retain them for auditing, recovery or selective undo (to be
explained later).


\begin{figure}[t]
    \centering
    \includegraphics[width=8cm]{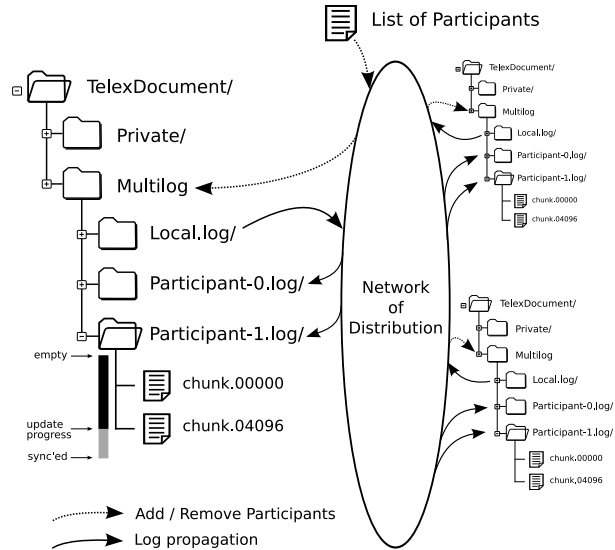}
    \caption{\label{fig:multilog-shared-document}
             Storage of Telex document}
\end{figure}

\section{Data structures}
\label{sec:structures}

\subsection{Document storage}


Telex stores its documents in file systems with standard,
best-effort consistency guarantees.
The storage design obeys some specific requirements.
Documents should be seamlessly integrated above a standard POSIX
interface, with reasonable performance and scalability.
They should co-exist with classical files and directories.
Participants must be able to work normally while disconnected.
The system should scales well with the number of collaborating
participants.
Finally, Participants' data must be secured even when shared.

We implemented multilogs above the federative peer-to-peer file system
VOFS \cite{DBLP:conf/grid/ChazapisTVKSK07}.
VOFS provides global access to files with best-effort consistency.
It supports disconnected operations via persistent replication,
and notifications for file modifications on distributed files.
A complete description of VOFS is outside of the scope of this paper;
here we focus on specific features related to Telex integration.

\subsubsection{Multilog Design}


As illustrated in Figure~\ref{fig:multilog-shared-document},
a Telex document is a structured directory of files.
Applications and Telex may store document-specific data
within the document, such as filters and snapshots.
These data are local to a participant; only the multilog needs to be
replicated.

A multilog is itself structured as a directory that contains an
append-only log per participant.
Actions and constraints created by an application are appended to that
participant's log.
Each participant's log is replicated at the other participants' sites;
VOFS propagates the updates to the network.
As each log has a single writer, is append-only, and local to a
document, this avoids write contention and scalability issues.

Propagation of a log through the network is asynchronous, i.e., a log
replica may contain only a prefix of its source, as indicated by the
``sync'' bar in the figure.
Telex instances monitor the logs for new updates.
Eventually, all actions and constraints are known to all participants.

As time passes, an action eventually becomes committed and
is not needed any more.
To enable removing such old records,
a log is itself structured as a directory of chunk files.
When the size of the current chunk reaches a threshold,
a new one is created.
The name of a chunk file includes a sequence number, making it
convenient to read chunks in order, and to selectively delete chunks.
A chunk may be deleted when all the actions it contains are committed
and there is a later materialised snapshot.
This is, however, a policy decision;
a site may decide instead to retain old chunks for auditing or recovery.

\begin{figure}[t]
    \centering
    \includegraphics[width=8cm]{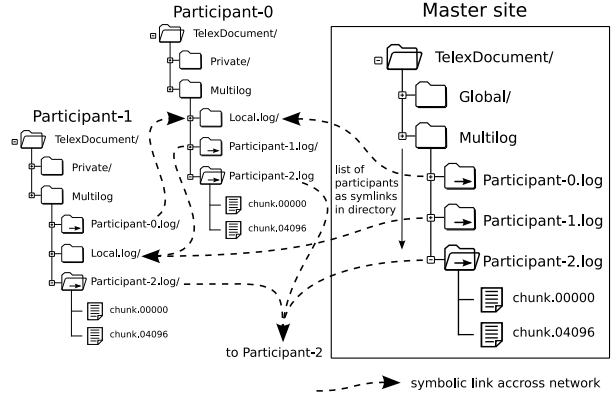}
    \caption{\label{fig:dfs_multilog_p2p}
             Implementation of multilogs over VOFS.}
\end{figure}

\subsubsection{Multilogs on VOFS}

A document is stored by the Telex \daemon in the file system as a directory.
The internal structure of this directory is not meaningful to users, and
is intended to be hidden by the user interface (much like the ``bundles''
of MacOS).

In our deployed multilogs so far, we have used a centralised setup at a primary
master site, containing the authoritative version of all the logs in a
document.
Participants' sites cache the logs persistently, making them available
for disconnected operation.
The master site is a single point of failure and a scalability
bottleneck.

In the future, we plan to use a peer-to-peer configuration,
using accross-network symbolic links that VOFS provides.
Here, each participant hosts the authoritative version of his own log on
his own site, as in Figure~\ref{fig:dfs_multilog_p2p}.
As before, participants cache remote logs persistently.
The master site serves only to list all the logs using symbolic links.
Any other method of distributing the list could be used.

\subsubsection{The Multilog Toolkit}

VOFS is optimised for multilogs, which improves the user experience.
However, multilogs can be implemented above any ordinary distributed
file system.
We provide a toolkit implementation of multilogs, as a set of simple
programs and d{\ae}mons, providing simple and efficient multilog
management and access above an ordinary file system.

The implementation follows closely the design of
Figure~\ref{fig:multilog-shared-document}.
More details are available in Section~\ref{sec:evaluation}.

\subsection{Action and Constraint}

\begin{figure}[t] \centering
  \includegraphics[width=.9\columnwidth]{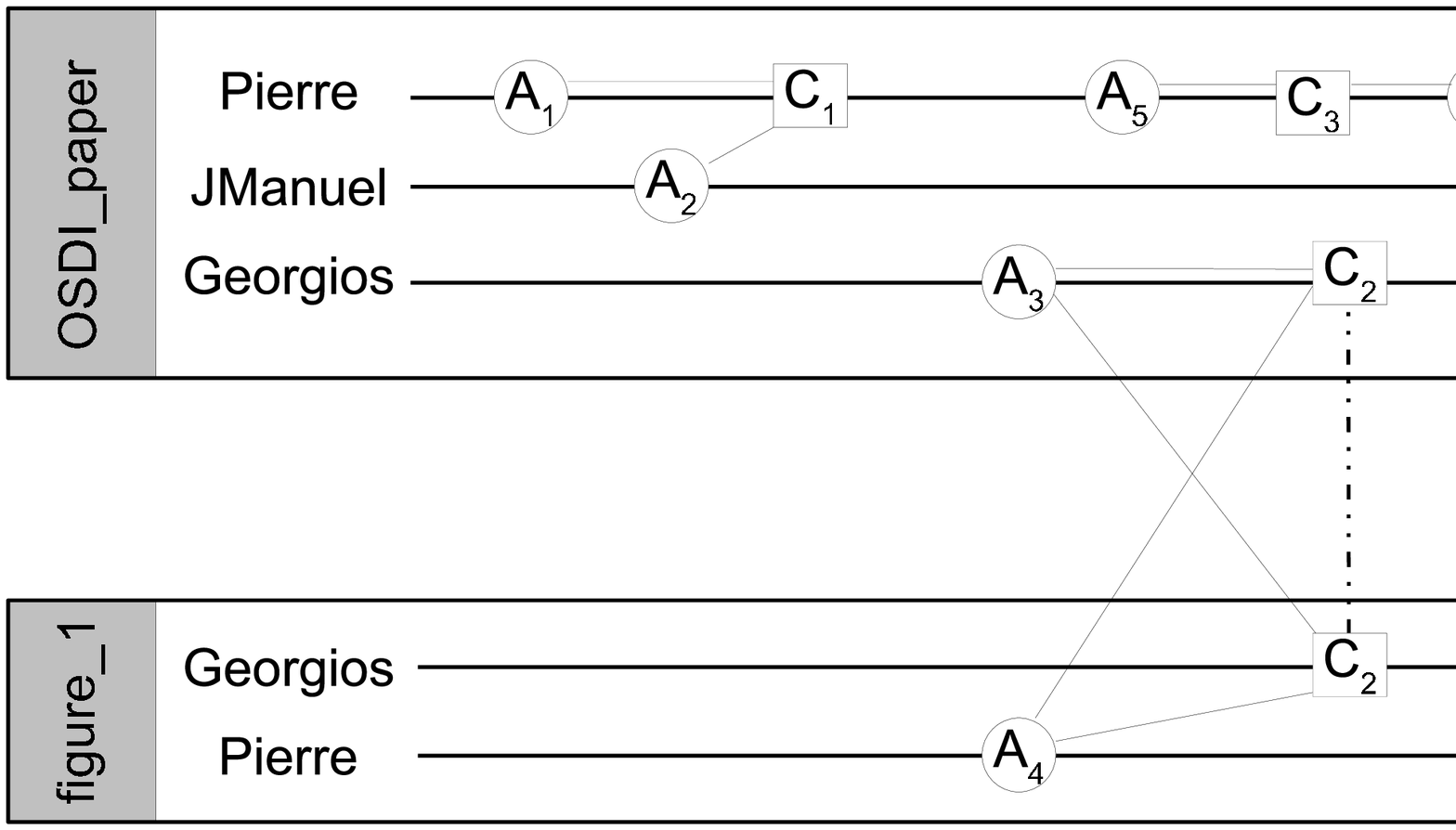}
  \caption{Two multilogs with their logs; note constraints within log,
    within document, and between documents}
  \label{fig:constraintLogging}
\end{figure}

An action represents an application operation. 
It is described by several attributes, of which some are known to Telex and
other are application-specific.
Among the former, the most important is a list of \emph{action keys}.
An action key indicates the document subset that this action
targets; if two actions have a common key, this indicates suspicion
that the actions conflict (see Section~\ref{sec:crossSiteConst} for more
detail).
An action belongs to only one document.
It is uniquely identified by the triple $\langle \tit{document, issuer,
  timestamp} \rangle$.
Telex logs an action in the log of the participant who issues it.

A constraint reifies a semantic relation between two actions.
It is defined by its type (\NonCommutingTxt, \NotAfterTxt or \EnablesTxt) and
by the two actions it binds.
A constraint is uniquely identified by the triple $\langle \tit{type, action1, 
action2} \rangle$.
Telex logs a constraint in the log of the participant who issues it.

Most often, a constraint binds two actions of the same document, whether
issued by the same participant or not.
Such a constraint is called an \emph{intra-document}
constraint. However, a constraint may bind actions of two distinct
documents.
Such a constraint is called a \emph{cross-document} constraint. It is
then logged in \emph{both} documents.

A constraint \tit{C} references an action \tit{A} by using one of the
three following forms: \tit{(timestamp)} if \tit{A} is issued by the
same participant as \tit{C} and belongs to the same document, \tit{(issuer,
timestamp)} if \tit{A} belongs to the same document as \tit{C} and
\tit{(docId, issuer, timestamp)} otherwise.
In the latter form, \tit{docId} is the id of the document that
action \tit{A} belongs to.

Figure~\ref{fig:constraintLogging} shows an example of the two types of
constraint.
Constraint $C_1$ is an intra-document constraint: it binds actions
$A_1$ and $A_2$ of document \tit{OSDI\_paper}.
Constraint $C_1$ is issued by \tit{Pierre} and thus it is logged
in \tit{Pierre}'s log of \tit{OSDI\_paper}.
On the other hand, constraint $C_2$ is a cross-document
constraint: it binds action $A_3$ of document \tit{OSDI\_paper} and action
$A_4$ of document \tit{figure\_1}.
Constraint $C_2$ is issued by \tit{Georgios} and thus it is logged
in \tit{Georgios}'s log of both \tit{OSDI\_paper} and
\tit{figure\_1}.

\subsection{Views}

A desirable feature of replication in collaborative work is to enable
different participants to have their own view of a shared document.
For instance a participant working on a given section of a shared document
 may temporarily ignore updates to the same section by other participants.
Telex allows the participant to select a particular view of a document by means of \emph{action filters}.
A filter defines which actions of the ACG Telex must
exclude when computing sound schedules. 
When applying a filter, Telex also exclude all actions that filtered 
actions enable.
This ensures that the view computed by filtering is always sound, i.e.,
document invariants are not violated.

A participant defines a filter by specifying its \emph{name} and one or more
filtering \emph{criteria} involving any  attribute of an action.
The participant may define several filters on a document and dynamically add
and remove them.
Telex saves currently-defined filters as part of the persistent state of a document.

Note that a filter may target a specific action of a document.
By adding and removing the filter, user may thus selectively
undo and redo the corresponding action in his view of the document.
(To undo an action persistently, the participant must \emph{abort} it.
By convention, this is expressed by marking the action as antagonistic
with itself.)

Filters also provide a means to permanently exclude the operations of a
 participant who turns out to be malicious, as in the Ivy file
  system~\cite{fic:rep:1611}.
Contrary to Ivy, Telex filters maintain correctness, by excluding all actions
that depends on the malicious participant's actions.

\subsection{Snapshot}

A snapshot records some view of the document.
To define a snapshot, a participant specifies its \emph{name} and the
\emph{schedule} of actions whose execution yields the state being
recorded.
In addition, the application may provide the corresponding binary state
of the document.
In this case, the snapshot is said \emph{materialised}.
Materialised snapshots speed up the computation of a view and are used
as garbage collection points.

The participant may define any number of snapshots of interest to him, and
later remove those that are no longer useful.
Telex saves the set of currently-defined snapshots as part the
persistent state of the document.



\section{Telex architecture and operation}
\label{sec:architecture}

\begin{figure}[t] \centering
  \includegraphics[width=.9\columnwidth]{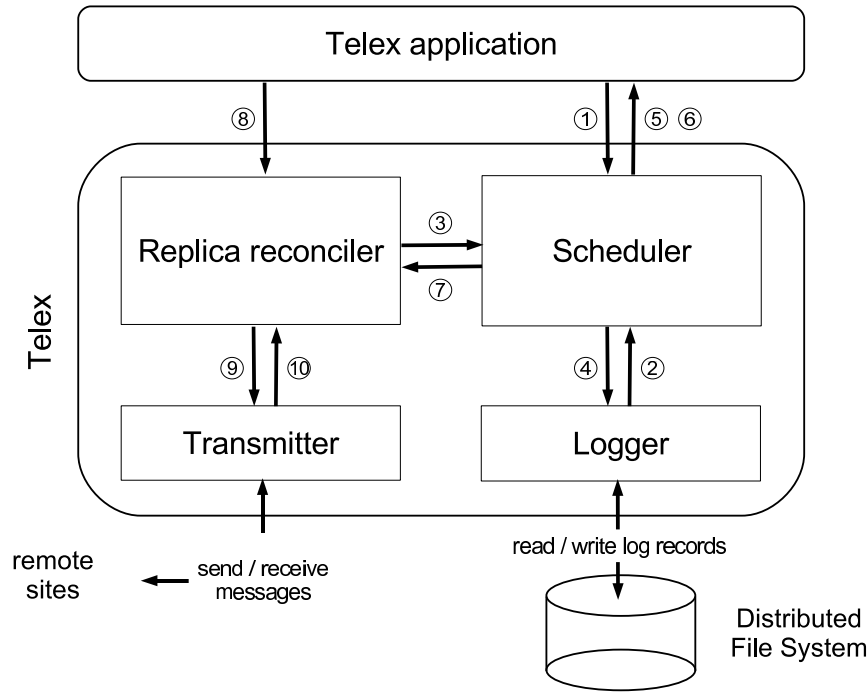}
  \caption{Telex architecture}
  \label{fig:telexArchitecture}
\end{figure}

Figure~\ref{fig:telexArchitecture} is a detailed view of
Figure~\ref{fig:app-scenario} which shows the overall architecture of Telex. 
An instance of Telex runs at each site and communicates with remote sites.

On top of the figure are the applications using the services of Telex.
Several such applications may run concurrently at the same site.
In the middle of the figure is the Telex system.
It is composed of two main modules --- the scheduler and the replica
reconciler --- layered on top of two auxiliary modules --- the
transmitter and the logger.
Arrows in the figure represent invocation paths between Telex modules
and to/from applications.

Each application may open one or more documents.
For each open document, Telex creates one instance of each module, which
maintains the execution context of the document.
The only exception is when documents are bound by cross-document
constraints, as described in section~\ref{sec:boundDocument}.
In this case, the bound documents share the same instance of the replica
 reconciler and the scheduler.


We describe next the interaction between a Telex instance and the outside world and then detail the operation of the main modules. 


\subsection{Interactions}
\label{sec:api}
Telex-application interactions involve exchanging pieces of AC graphs
(sets of actions and constraints downwards, sets of schedules upwards).
The interaction cycle is as follows. 
The participant acts upon the application, which translates his request into
one or more actions and constraints and passes them to Telex.
In return, Telex computes a sound schedule from the set of locally-known
 actions and constraints and hands the schedule to the application.
The application executes the schedule and presents the resulting state to
the participant.
If some actions conflict, then several sounds schedules exist, each 
corresponding to a possible solution to the conflict.
The application presents the resulting states to the participant so that he
can select the solution he prefers.

Telex sites exchange actions and constraints through multilogs, and
communicate with each other in the commitment protocol.
The logger module logs the actions and constraints submitted by the local 
participant in the participant's log.
In return, the VOFS notifies the logger when remote participant's log are updated.
The transmitter determines the set of peer sites and provides an Atomic Multicast service among peer sites (arrows~\#9
and \#10).


%
%
%
%

\subsection{Scheduler}
\label{sec:scheduler}


The role of the scheduler is twofold.
First, it maintains the in-memory ACG that represents the state of the document
 at the local site.
Second, it periodically computes sets of sound schedules from the ACG 
and proposes them to the application for execution.
Actions and/or constraints are added to the graph either by:
\begin{itemize}
\item The application (Figure~\ref{fig:telexArchitecture}, arrow~\#1),
  when the local participant updates the document.
\item The logger (arrow~\#2), when it receives an update issued by a
  remote participant.
\item The replica reconciler (arrow \#3), when it commits a schedule.
\end{itemize}

The scheduler passes locally-submitted actions and constraints to the
logger (arrow~\#4) to log them on persistent storage.

\subsubsection{Cross-site constraint generation}
\label{sec:crossSiteConst}

Actions logged independently by two participants may conflict; for
instance in the shared calendar application, a same user could be added
to two parallel meetings.
Telex ensures that conflicts are reified by constraints as follows.
When a site receives a new action, it compares it against already-known,
concurrent actions of the same document.
If they have a common key, then Telex invokes the corresponding
application's \tit{getConstraint} upcall.
If the actions really conflict, the application responds by logging an
appropriate constraint (arrow~\#5 in Figure~\ref{fig:app-scenario}.b or
Figure~\ref{fig:telexArchitecture}).

Action keys are opaque to Telex, which tests them for equality only.
Action keys serve as a compact, but approximate, representation of the
document subset that the action uses or updates.
Typically, an action key hashes the identifier of a parameter of the
action.
Multiple keys have ``or'' semantics (Telex upcalls \tit{getConstraint}
if a key of one action equals any key of the other).
To implement ``and'' semantics (for instance, to get an upcall only if
two given objects are involved) the application hashes the XOR of
their identifiers into a single key.
An action with no keys conflicts with no other.

If two unrelated actions happens to have equal action keys, no harm is
done, other than a loss of performance.

\subsubsection{Schedule generation}

A large number of sound schedules exist for any given ACG in the general case.
It is therefore not feasible to compute all sound schedules beforehand
and present them to the application.
Besides, the application may be interested only in a few or even just
one schedule.
For these reasons, Telex generates sound schedules dynamically, upon
application request (this is not shown in Figure~\ref{fig:telexArchitecture}).
The application may thus iterate through the proposed schedules and
stops when one or more appropriate schedules are found.

Telex generates the best schedules first, where the quality metric is
the number of actions included (implying fewer actions aborted).
Optimal scheduling is NP-complete, therefore Telex runs a heuristic
inspired by IceCube \cite{optim:syn:1474}.
Secondary goals of the heuristic are to give preference to actions of
the local participant in the case of a conflict, and to avoid returning
a schedule equivalent to one returned previously.

%
%

\subsubsection{Bound documents}
\label{sec:boundDocument}

Two documents are said bound if there exists a constraint between an
action of one and an action of the other, and either action (or both) is
not committed.
For instance, if a participant wishes to update two documents atomically, he
sets an \EnablesTxt constraint in each direction between the updates.

The actions of a document may not be scheduled independently from those
of the documents it is bound to.
Scheduling is optimised for the common case of non-bound documents, but
we provide special processing for this particular case.
Note that bound documents may be handled by distinct applications.

Telex processes bound document by merging them into a single \emph{shared} 
ACG in order to compute \emph{global} schedules over all actions and constraints.
Each global schedule generally contains actions from all bound
documents.
Thus, in order to execute a global schedule, Telex first projects
the schedule on each document and passes each resulting sub-schedule
to the relevant application.
The projection operation simply consists in retaining only those actions
that belong to the target document while preserving their order.
Telex assigns the same identifier to the sub-schedules deriving from the
same global schedule.
This way, the participant can identify matching sub-schedules on each bound
document.

\subsection{Replica reconciler}
\label{section:reconciliation}
Each Telex site proposes a set of constraints, a \textit{proposal}, to remote sites.
A proposal contains decision to commit, abort or serialise actions.
These proposals may differ, due to asynchronous communication,
filtering, differing local information, or user preference.
The \emph{replica reconciler} is in charge of \emph{commitment}, i.e.,
reaching agreement on a common schedule prefix.
Commitment occurs in the background, not within the critical path of
applications.
The committed proposal appears as a prefix of the local schedules.

We propose a plug-in replica reconciler architecture, providing different
strategies according to needs.
A reconciler has four (asynchronous) phases.

\begin{enumerate}
\item Each sites compute a proposal, according to its local view, for
  instance based on the user's preferences (arrow~\#8 in Figure~\ref{fig:telexArchitecture}).
\item The transmitter atomic multicasts proposals to set of sites
  directly concerned (arrow~\#9) by the agreement (in case of bound documents
  more than one replica group may be concerned).
  Atomic multicast maintains liveness in presence of faults and network lags.
\item The transmitter forwards proposals it receives up to the replica
  reconciler (arrow \#10).
\item According to the commitment algorithm (described next) the
  reconciler chooses a winning proposal, and logs it (arrows \#3 and \#4).
\end{enumerate}

\noindent
Currently we propose two commitment algorithms.
(i) A first-in first-out algorithm for applications such as a
  distributed database.
  At each site the FIFO algorithm proposes to minimise the number of dead actions
  according to its local view.
  When a site delivers a new proposal, the FIFO algorithm checks
  the soundness of the proposal according to the previous winning
  proposals (arrows \#8 and \#7).
  If the decision is sound, the reconciler adds it to the ACG,
  if not the decision is discarded.

(ii) A voting algorithm that takes into account local preferences.
  A proposal is a vote spanning one or multiple actions over one or more documents.
  A proposal is broken into sub-ACGs with specific properties, called candidates.
  Candidates containing the same actions challenge each other.
  A candidate may be elected only if its set of actions is transitively
  closed in the union of all the ACGs across sites.
  This protocol is described in detail in a separate publication
  \cite{rep:1592}.

%


\subsection{Access control}

The Telex design includes access control at increasingly fine-grain
levels, using a security framework (whose description is out of scope of
this document).
This is indicated by the three arrows marked \textsf{check} in
Figure~\ref{fig:app-scenario}.
(i) Access control at file granularity ensures that a single participant
writes a given log, and that only authorised users can read a log.
(ii) The Telex d{\ae}mon checks whether a user is allowed to access an individual
log record.%
\footnote{
    This is not yet implemented in the current version.
}
(iii) Applications may enforce further control.
For instance, in the SC application, a user might observe the times that
another user is busy, but not be allowed to see the other details of his
meetings.
As explained in Section~\ref{sec:engineering}, access control does not
violate consistency.

\section{Applications}
\label{sec:applications}

To provide insight on the issues involved in using the Telex system, 
this section presents some of our example applications.
We will return to the lessons learned in a later section.

\subsection{Simple Replicated Dictionary}


\newcommand{\Attribute}{\ensuremath{\mathit{attribute}}}
\newcommand{\Attrs}{\ensuremath{\mathit{attrs}}}
\newcommand{\Attr}{\ensuremath{\mathit{attr}}}
\newcommand{\Insert}{\ensuremath{\mathit{insert}}}
\newcommand{\Ins}{\ensuremath{\mathit{ins}}}
\newcommand{\TupleId}{\ensuremath{\mathit{tupleID}}}
\newcommand{\TID}{\ensuremath{\mathit{TID}}}
\newcommand{\TIDs}{\ensuremath{\mathit{TIDs}}}
\newcommand{\Modify}{\ensuremath{\mathit{modify}}}
\newcommand{\Mod}{\ensuremath{\mathit{mod}}}
\newcommand{\Read}{\ensuremath{\mathit{read}}}
\newcommand{\Remove}{\ensuremath{\mathit{remove}}}
\newcommand{\Rem}{\ensuremath{\mathit{rem}}}
\newcommand{\Set}{\ensuremath{\mathit{set}}}
We start with a simple example.
Our Simple Replicated Dictionary Application (SRDA) manages shared
dictionaries.
SRDA is intended as a building block for applications such as a shared
address book.
Users can operate on a dictionary in either connected or disconnected mode.
Telex guarantees that, in spite of node arrivals, departures or
failures, all instances of a given dictionary converge. 

A document contains tuples of the form $\langle \TupleId, \Attribute_1,
\Attribute_2, \ldots \rangle$, for any number of attributes.
Each attribute is a  $\langle \textit{name, value} \rangle$ pair.
SRDA provides these operations:
\begin{itemize}
\item $\Insert (\TupleId, \Attrs)$: inserts a new entry, with identifier
  \TupleId and attributes \Attrs, into the
  dictionary document.
\item $\Modify(\TupleId, \Attrs)$: modifies attributes for
  the given \TupleId.
\item $\Remove (\TupleId)$: deletes the tuple corresponding to the
  given \TupleId.
\item $\Read (\TupleId)$: returns the attributes corresponding to the
  given \TupleId.
\end{itemize}

In the first operation, the \TupleId must be previously unused or removed; for all
the others, a tuple identified by \TupleId must already exist.
The modify operation assigns the listed attributes if they already exist
for the tuple, otherwise it adds them.

Insert, modify and remove operations translate to a Telex action.
Because Telex does not yet support isolated multi-operation
transactions, we manage write dependencies in the write operations, as
explained shortly.
Read operations are treated as local.

\begin{table}[t]
   \centering
   \begin{tabular}{@{}c|l@{}}\scriptsize
     \Insert & $\forall \mathrm{~previous~} \Rem_i.\TID:$ \\
     				 & \qquad $\Rem_i.\TID$ \NotAfter current $\Ins.\TID$ \\
     \hline
     \Remove & $\Ins.\TID$ \Causal current $\Rem.\TID$
\\
     \hline
     \Modify & $\Ins.\TID$ \Causal current $\Mod.\TID$ \\
             & $\forall \mathrm{~previous~} \Mod_i.\TID.\Attr_j:$ \\
             & \qquad $\Mod_i$ \NotAfter current $\Mod$
   \end{tabular}
   \caption{Sequential execution constraints (Notation: \Ins = \Insert, \Mod =
     \Modify, \Rem = \Remove, \Attr = \Attribute, \TID = \TupleId)}  
   \label{tab:SRDAconstraintsActions}
\end{table}

\subsubsection{Sequential constraints}

Table~\ref{tab:SRDAconstraintsActions} summarises the sequential
semantics of SRDA.
SRDA logs these constraints at the same time as it logs the right-hand
action of the constraint.

In the Telex design, the application should log causal dependence only
when the second action truly depends on the first.
Hence, a \Modify action, or a \Remove, is causally dependent on the
\Insert that created the tuple.
Thus, if the \Insert aborts or fails, the dependent \Modify and
\Remove actions will be discarded from any sound schedule. 
Furthermore, we treat every write operation as a read-compute-write
transaction.

In order to ensure read-your-writes session guarantees
\cite{rep:syn:1481}, we set \NotAfterTxt constraints between \Insert,
\Modify and \Remove actions in the same user session, even between
different dictionary documents.

Finally, to ensure the correct scheduling of a \Remove followed by an
\Insert with the same tuple identifier, we make all
previous \Remove with the same tuple-id \NotAfterTxt the current \Insert.
The SRDA application logs the above constraints in the multilog, at the
same time as it logs the right-hand action.

The SRDA application logs the above constraints in the multilog, at the
same time as it logs the right-hand action.

\begin{table}[t]
   \centering\scriptsize
   \begin{tabular}{@{}c|c|c@{}}
                 &      $\Ins_2$                          &     $\Mod_2$                                \\
     \hline
     $\Ins_1$    & $\Ins_1.\TID=\Ins_2.\TID$              &                                                \\
     		 & $ \implies \Ins_1 \NonCommuting \Ins_2$  &                                  --            \\
     \hline
                 &                                        & $\Mod_1.\TID=\Mod_2.\TID \land$          \\
     $\Mod_1$    & impossible                             & $ \Attrs_1.\TIDs \inter
                                                                     \Attrs_2.\TIDs \neq \emptyset$              \\ 
                 &                                        & $ \implies \Mod_1 \NonCommuting \Mod_2$  \\
   \end{tabular}
   \caption{SRDA \tit{getConstraint}}
   \label{tab:SRDAconstraints}
\end{table}

\subsubsection{Concurrency constraints}

Since it is illegal to insert the same identifier twice, two concurrent \Insert
actions that refer to the same identifier are \NonCommutingTxt.
Otherwise, concurrent inserts commute.
Similarly, two concurrent \Modify operations with the same identifier and overlapping
attributes are also \NonCommutingTxt.

Those constraints are added by the application when Telex invokes its
\tit{getConstraint} method.
They are summarised in Table~\ref{tab:SRDAconstraints}, where
\NonCommutingTxt is noted \NonCommuting.
In order to ensure that Telex upcalls the \tit{getConstraint} method
as needed, \Insert and \Modify actions have an action key, computed as a
hash of the \TupleId.

\subsection{Shared Calendar}

Our Shared Calendar (SC) application is representative of collaborative
decision-making applications.
SC illustrates the advantages of Telex for semantically-rich
collaborative applications.

SC helps people organise private events and group meetings
collaboratively, possibly in disconnected and asynchronous mode.
Contrary to existing calendar applications, SC detects conflicts (such
as double booking), proposes solutions, and ensures agreement and
eventual consistency.

This would be difficult to achieve without Telex support.
Application logic (i.e., maintaining the data
structures and identifying constraints) is well separated from the
system logic, i.e., persistence, replication, conflict detection and
resolution, commitment, etc.

\subsubsection{SC logic}
\label{para:SCLogic}
Each user or location has an associated \emph{calendar} document.
Each \emph{event} (e.g., a meeting) is a separate document.
A calendar may be read or updated by other users, who can (if so
authorised) create or manage events, invite people to an event, or
identify conflicts and free time.

We use the following notations.
An event \tit{e} is unique, 
has a name \tit{e.name}, and a date\tit {e.date},
and is materialised by a Telex document \tit{e.dox}.

A user \tit{A} creates an event \tit{e} by creating the document
\tit{e.dox}, and by logging an \tit{open-event} action in his own
calendar and an \tit{invite(A)} actions in \tit{e.dox}.
He also logs an \tit{enable-event} action in \tit{e.dox} that symbolises
the creation of the event.
This action is used to specify constraints on the event creation as
shown next.


Later, user \tit{A} may invite other users by logging an
\tit{open-event} action in his log within their calendars, and a
corresponding \tit{invite} action in \tit{e.dox}.

Once a user has opened an event document, he may invite more users.
He also can cancel the event or some user invitation by logging a \tit{cancel-event} or a \tit{cancel-invitation}  action in \tit{e.dox}.%
\footnote{
    Currently it is not possible to collaboratively change the time of an event.
    This will require extensions to Telex to associate the time updates with some user invitation to detect a double booking, which is future work.
}
%

The action keys identify the event and its time-slots.
Therefore, actions in the same calendar for the same event, or for
different events at the same time, will have overlapping keys, causing
Telex to invoke the \tit{getConstraint} upcall interface of SC.

A calendar document action commutes with all other calendar document
actions.
Constraints between event document actions are similar to the SRDA
constraints, where \tit{enable-event}, \tit{cancel-event} and
\tit{invite} (or \tit{cancel-invitation}) are like
like  \Insert, \Remove and \Modify respectively.


To avoid double bookings, concurrent \tit{invite} actions are
antagonistic, if they concern the same user at the same time but
different events.

\begin{figure}[t]
    \centering
    \includegraphics [width=10cm]{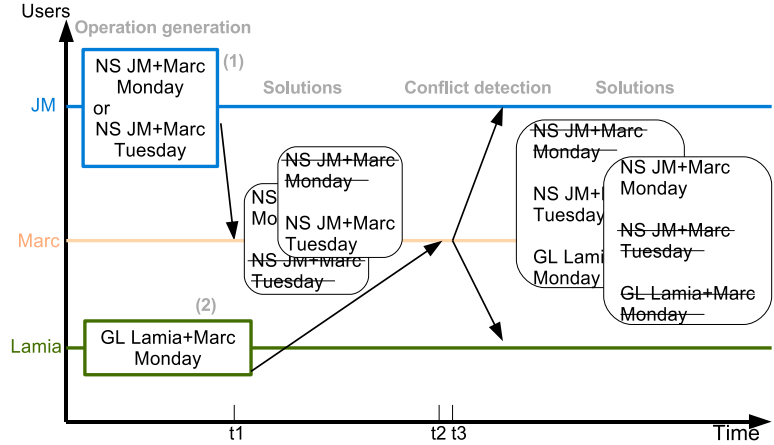}
    \caption{\label{ScenarioCalendarDiagram} Execution scenario for the  Shared Calendar application}
\end{figure}

\begin{figure}[t]
    \centering
    \includegraphics [height=8cm]{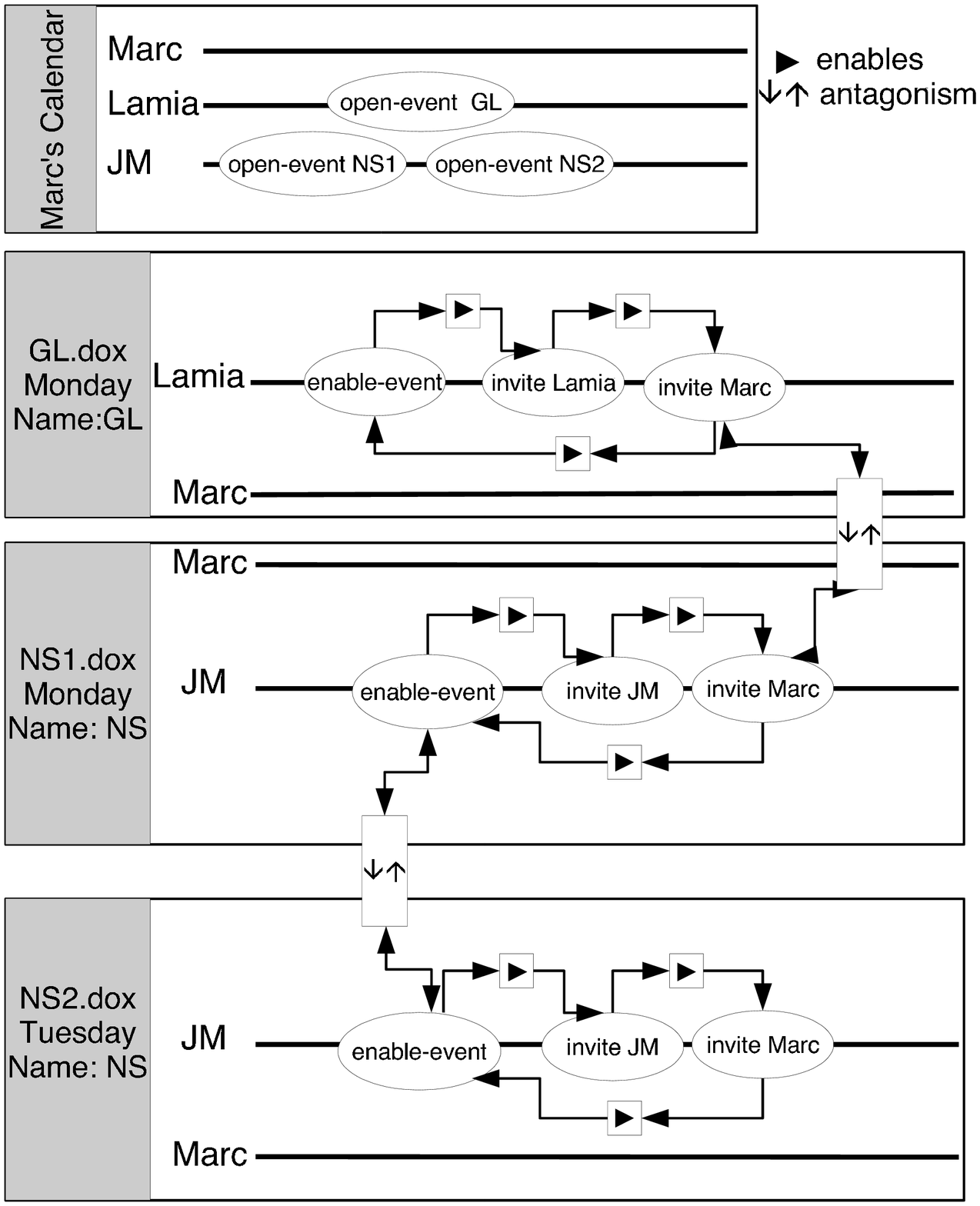}
    \caption{\label{MarcSite} Marc's site at $t3$}
\end{figure}

\subsubsection{Use case}

Consider the scenario in Figure~\ref{ScenarioCalendarDiagram}.
Users Jean-Michel, Lamia and Marc are working separately and communicate
only via the SC application.

 Jean-Michel organises meeting {Networking Seminar} \tit{NS} with Marc.
 He proposes two alternative dates, {Monday} and {Tuesday}
(Operation~1 in the figure).
 
Lamia also organises a meeting {Greek Lesson} \tit{GL} with Marc on {Monday}
(Operation~2). 


SC creates the event documents and logs the actions and constraints to
Telex, as detailed in Figure~\ref{MarcSite}, depicting the state of
Marc's site at time $t3$.


Lamia's SC instance creates \tit{GL.dox} document, imports Marc's calendar,
and logs the following actions:
\begin{itemize}
\item On Marc's and Lamia's calendar:  \tit{open-event (e2)}.
\item On \tit{GL.dox}: 
  $A=\tit{enable-event}, B = \tit{invite(Lamia)}, C = \tit{invite(Marc)}.$
  SC groups them atomically: $A \Atomic B \land B \Atomic C$.
\end{itemize}

To express the alternative Jean-Michel's SC instance transparently creates two events \tit{NS1} and \tit{NS2} with conflicting \tit{enable-event} actions.
For both events, SC generates similar actions as for the \tit{GL} event.

Suppose that, at some point in time $t1$, Marc has received
Jean-Michel's actions, but not yet Lamia's.
This may happen, for instance, if Lamia is working offline.
Telex computes the schedules corresponding to two possible solutions:
(i) holding \tit{NS} on {Monday} and aborting \tit{NS} on {Tuesday};  or
(ii) holding \tit{NS} on {Tuesday}, and aborting \tit{NS} on {Monday}.
Since the former solution contains more actions, it will be proposed
first.


Later, at $t2$ Marc knows Lamia's actions.
Telex checks the keys of Lamia's actions with Jean-Michel's.
$C = \tit{invite(Marc)}$ on \tit{GL.dox} and $E = \tit{invite(Marc)}$ on
\tit{NS1.dox} both have a key representing the \tit{Monday} slot.
Therefore, Telex asks SC for the corresponding constraints.
SC returns an antagonism constraint $C \Antagonism E$.
This ensures that no view contains both \tit{C} or \tit{E}, and that one
or the other (or both) eventually abort.


Finally, Telex offers the two possible solutions: (i) \tit{NS} on
{Tuesday} and \tit{GL} on {Monday}, aborting \tit{NS} on {Monday}; or
(ii) \tit{NS} on {Monday}, aborting \tit{GL} on {Monday} and aborting
\tit{NS} on {Tuesday}.


Lamia is not invited to event \tit{NS}, she may not read \tit{NS1.dox}
nor \tit{NS2.dox}.
Nevertheless, Telex ensures that she eventually gets notified of a
conflict occurrence that may abort \tit{GL}.
The same goes for Jean-Michel.
The reconciliation phase ensures that Marc, Lamia and Jean-Michel
eventually see a consistent state for \tit{GL} and \tit{NS} events.

\subsection{Shared wiki}

For lack of space, we describe our Shared Wiki Application (SWA) only
briefly.

Each wiki page is a separate document.
Every user currently editing it has a log in the document.
His site keeps a local replica of the wiki text, which the user modifies
locally using a standard text editor.
Every time the user saves, the SWA computes the difference from the
previous version, and translates it into insert-line and delete-line
actions.
Modifying a line is interpreted as an atomic grouping of delete-line and
insert-line.

The SWA uses the WOOTO operational transformation algorithm
\cite{app:rep:1587} to ensure that concurrent edit operations commute.
A delete-line action depends causally on the action that inserted the
line.
Inserting a line between two other lines depends causally on the two
corresponding line insert operations.

Since all concurrent operations inside a document commute, there will
never be any conflicts.
Therefore, edit actions carry no keys, and Telex never upcalls
\tit{getConstraint} to the SWA.
Schedule computation is trivial, since all schedules that are compatible
with causal dependence order are equivalent.

Existing wiki editors maintain the set of past versions of a page.
Thanks to Telex, SWA can reconstruct any past version, and additionally
maintains the relations between versions.
In the future, we could extract more history information from the
persistent multi-log, including page splits and merges, and copy-paste
between pages.

From the perspective a single page, Telex serves mainly to reliably
broadcast actions and replay them in causal order.
One added value of Telex for SWA is the ability to perform
multi-document updates, e.g., a global replace through all wiki pages
consistently.
Telex also enables multi-application scenarios, e.g., ensuring that a
wiki page contain the details of a meeting agreed in the shared calendar
application.

\section{Performance evaluation}
\label{sec:evaluation}

\begin{table}
\small
\begin{tabular}{r|rrrr}
  Config Name     &    1x8M  &   8x8M &    1x8L  &   8x8L  \\ \hline 
  Writers         &      1   &     8  &      1   &     8   \\ 
  Log size (MB)   &     50   &    50  &      5   &     5   \\ 
  RX limiting     &     no   &    no  &    yes   &   yes   \\ 
  runtime (sec)   &    3.4   &   9.3  &  306.48  & 309.31  \\ 
  avg RX+TX (B/s) &  102.9M  &  75.3M &  228.4K  & 226.3K  \\ 
\end{tabular}
\caption{Representative results for shared multilogs with
         1 and 8 writers, with and without
         limiting receiving traffic}
\label{tab:multilog_results}
\end{table}

\begin{figure}
    \centering
    \includegraphics[width=10cm]{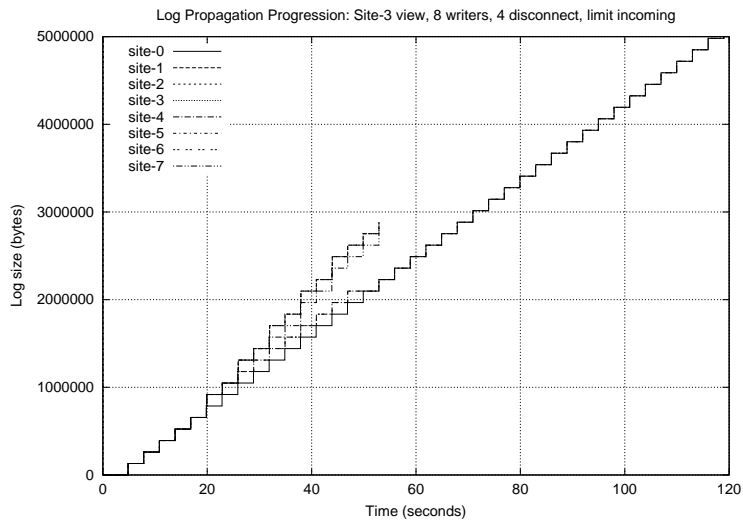}
    \caption{\label{fig:multilog-bench-limit-disc-8x8}
             Multilog replication progression for 8 writers,
             throttled incoming traffic. 4 disconnect.}
\end{figure}

\subsection{Multilog experiment}

The multilog toolkit is a simple set of tools and d\ae{}mons
that create, access and connect logs in multilogs.
It is written in Python and uses TCP/IP for networking.
It straightforwardly implements the design illustrated in
Figure~\ref{fig:multilog-shared-document}.

There are four main utilities in the toolkit.
\emph{LogServer} monitors a log and propagate updates.
\emph{LogClient} contacts a list of LogServers and locally
replicates their logs.
\emph{LogTool} is a utility that can read or write a log.
\emph{MultilogD} is a simple d\ae{}mon that given a list of participants,
combines the log-tools to implement a multilog.

\subsubsection{Evaluation summary}

The multilog structure decouples reads and writes and promotes
mostly-linear access patterns.
Therefore, the read{\slash}write performance of multilogs is dominated
by the local filesystem and of the network stack.
The purpose of this evaluation is to demonstrate this fact; the results
are summarised in Table~\ref{tab:multilog_results}

Our performance goals are to scale to very large numbers of readers.
The numbers of writers for a single document is expected to remain
relatively small, on the order of tens of participants.
This is typical for the internet society.

Efficient propagation from a small number of writers
to a huge number of readers is possible in peer-to-peer networks,
where recipients of data propagate them further.
The net effect of such a solution is a high outgoing bandwidth and
limited incoming bandwidth.
In some of our experiments, we emulate this effect by severely limiting
incoming traffic of participants while leaving outgoing traffic unlimited.

\subsubsection{Detailed Results}

The experimental setup involves one participant installed on
each of 8 nodes interconnected with Gigabit Ethernet.
The scenario is simple;
Either one or all 8 participants begin to log a specific amount of data as
fast as possible.
At the same time, each participant reads his logs and records its
replication progression over time.
The writers and readers are implemented with LogTool instances,
logs are served by LogServers
and propagated updates are received and written to replicas
by LogClients.

Table~\ref{tab:multilog_results} lists representative results
for running 1 and 8 concurrent writers both with and without
limiting the incoming traffic.
The average traffic is the sum of the incoming and outgoing
traffic combined.

Our conclusion is that, when there is no limit in effect,
multilog propagation performance is comparable
to the maximum network bandwidth.
When limits are in place, although overall bandwidth drops as expected,
we observe that varying the number of writers between 1 and 8 has no
effect.
Furthermore, in all the experiments, disconnection of a participant does
not disrupt the remaining ones,
as illustrated in Figure~\ref{fig:multilog-bench-limit-disc-8x8}.

\subsection{Synthetic benchmarks}

\paragraph{Sound schedules computation} Telex computes sound schedules
using the IceCube algorithm \cite{optim:syn:1474}.
For a randomly generated graph containning 10000 actions and 20000 constraints,
our algorithm computes a sound schedule in 200 ms.
In running mode Telex uses incremental mode,
and the computation is around a millisecond.

\paragraph{Reconciliation time}
We test the time to decide newly proposed actions.
During this experience we compute a schedule every 100ms,
and a proposal every 100ms.
Each site submits 20 actions per second.
The average time to commit an action using 
the FIFO algorithm (see Section \ref{section:reconciliation})
is 64ms.

\subsection{STMBench}

We run the STMBench7 benchmark \cite{DBLP:conf/eurosys/GuerraouiKV07}, which
emulates an application with a rich data structure and many different
operations.
We chose STMBench7 mainly because it demonstrates concurrency and
conflicts.
It also serves as an illustration of the use of Telex on a complex data structure.

STMBench7 was developed to exercise software transactional memories,
based on the previous OO7 benchmark for object-oriented databases.
STMbench7 builds an object graph with millions of objects and connected
by numerous pointers.
It contains 45 operations (21 read-only, 24 read-write) with various
scope and complexity.
We ported to Telex the read-write operations only.
They all operate in a similar manner: traverse the data structure,
reading one or many attributes of one or many objects, and modify an object.

An STMBench7 benchmark consists of two phases: creating a randomised
object graph, and invoking operations.
We measure only the second phase.
There are four four main categories of operations:
\begin{itemize}
\item Long traversal: access large parts of the object graph, typically
  all ``assemblies'' and ``atomic parts''.
\item Short traversals: access fewer objects, traversing the graph along
  a randomly chosen path.
\item Short operations: choose a small number of objects, and perform an
  operation on these objects or in their neighbourhood.
\item Structure modifications: randomly create or delete objects, or create or
  delete pointers between objects.
\end{itemize}


Each STMBench7 operations is mapped to a single action, hence will
be isolated from concurrent operations.

Unexpectedly, in the original code, operations always commute, because
the updates either swap two shared pointers, or add 1 modulo 2 to a
shared integer.
We therefore modified the benchmark so that, with some probability,
updates either commute or do not commute.

Due to the large number of operations, we will not present a
comprehensive list of constraints.
Instead, we explain the rules we follow to define the constraints.

\begin{itemize}
\item Any modification to an object is causally dependent on the
  creation of the same object.
\item Two actions that modify the same data are  \NonCommutingTxt.
\item If an action reads some data, and another action concurrent writes
  the same data, the former is \NotAfterTxt the latter.
  This ensures that, at all sites, the read will see the value before
  the write.
\end{itemize}

The results of the benchmark are shown in Table~\ref{tab:stm}, executing
the operations that modify data (not the structure).
Performance is independent of the number of sites.

\begin{table}
  \centering
  \begin{tabular}{cc}
    Number of sites       & Time to benchmark (s) \\
\hline
    1                     & 20                     \\
    2                     & 21                     \\
    3                     & 21                     \\
    4                     & 21                     \\
    5                     & 21                     \\
    6                     & 21                     \\
  \end{tabular}
  \caption{STMBench7 results}
  \label{tab:stm}
\end{table}

\section{Lessons learned}
\label{sec:lessons}

Experience with applications and benchmarks has given us useful
feedback, both regarding the implementation of Telex, as well as
guidelines for application developers.

The current implementation of Telex suffers from excessive memory
consumption.
The ACG can quickly reach sizes of several tens of thousands of nodes, and
is accessed concurrently by many threads.
For instance, the scheduler parses the ACG at the same time as local and
remote applications are modifying it.
To avoid concurrency issues, the scheduler takes a full copy of the current
ACG, which both consumes memory and is slow (in Java).
Similarly, forward execution and rollback of applications involves
copying their internal state, which can be very large.
In both cases, an obvious solution (and future work) is to copy-on-write
instead.

Translating application semantics into actions and constraints is a
skill that takes time to acquire.
We present some guidelines derived from our own experience.
Note that these are not hard rules, and even may be conflicting.

The most important suggestion is to leverage commutativity
as much as possible.
As noted in the SWA, if all operations commute, consistency is trivial.
The SWA example also shows that, sometimes, operations that appear
non-commuting intuitively, can be designed or transformed to commute.

We learned that it is important to turn every piece
of shared information into a separate document.
In the initial design of SC, calendars were the only documents, and
events were implicit in the calendars.
This raised a number of problems, because there was no obvious way to
detect when a meeting conflict would impact another user indirectly.
Separating out events as distinct documents solved this.

It is important to distinguish the sequential constraints (mainly,
\NotAfterTxt and \CausalTxt) from the concurrency constraints
(conflicts).
The former are logged with their right-hand action; the latter are
logged in response to \tit{getConstraint}.
Concurrency constraints are derived from the application invariants.
For instance, in SRDA, the sequential specification forbids two tuples
with the same identifier; it follows that concurrent \Insert{}s with the
same identifier are in \AntagonismTxt.

One lesson from STMBench7 is to reason about high-level operations
rather than low-level ones, in order to deal with fewer combinations.
Furthermore, it is sometimes the case where high-level operations
commute (for instance, increment and decrement a shared integer) even
though their low-level implementations (e.g., reads and writes) do not.

However, in some cases, it may be simpler to reason about a small number
of low-level primitives when they may be combined into a large number of
operations.
Currently, this kind of approach is complicated by the lack of support
for transactional isolation, which is future work.

Constraints are hard to validate.
We suggest two complementary approaches for future work.
A compiler could generate actions and constraints from a high-level
specification, and a checker could verify that all action-constraint
combinations verify the application invariants.

\section{Related work}
\label{sec:related}

State-machine replication \cite{con:rep:615} is based a total order of
operations.
This ensures consistency and correctness, but requires consensus at each
operation, in the critical path of the application.
In contrast, Telex's optimistic approach performs consensus in batches,
in the background.

Optimistic replication \cite{optim:rep:syn:1500} has been widely used,
e.g., in replicated file systems (for instance, Coda \cite{fic:rep:887}
or Roam \cite{fic:rep:syn:optim:1445}) and for collaborative work (e.g.,
Bayou \cite{syn:optim:rep:1433}).
In these systems, replicas eventually converge, but they generally do
not ensure any high-level correctness.
For instance, the widely-used ``last-writer-wins'' (LWW) loses updates
when conflicts occur, and does not maintain consistency between objects.
Our constraints additionally ensure that application invariants are
preserved.

Many replicated systems transmit new values or deltas (the state-based
model).
The operation-based model used in Telex (i.e., the system stores,
transmits and replays logs of operations) retains more useful
information for reconciliation.
This is especially advantageous when high-level operations logically
commute despite reading and writing the same physical data, as in our SC
and SWA applications.

The literature on computer-supported co-operative work is widely based
on operational transformation (OT) \cite{alg:rep:1460}.
OT ensuring commutativity between concurrent operations by modifying
them at replay time.
Combined with reliable causal-order broadcast, this ensures convergence
with no further concurrency control, but unfortunately OT appears
limited to very simple text-editing scenarios.
Telex takes advantage of commutativity when it is available, and
supports any mix of commutative and non-commutative operations.

Coda's application-specific resolvers \cite{fic:rep:syn:1290} or Bayou
\cite{syn:optim:rep:1433} give applications full control over conflicts.
However, this requires developers to have a deep understanding of
distributed systems issues.
Instead, Telex requires stylised concurrency constraints from
applications and takes care of conflict resolution in an
application-indendent manner.


Telex has many similarities with Bayou \cite{syn:optim:rep:1433} and
also many differences.
Bayou is an operation-based system that provides commitment; the
committed state is guaranteed correct.
However, Bayou relies on a primary site for commitment and the committed
schedule is unpredictable.
Furthermore, the system offers no help for reconciliation.

Constraints were used for reconciliation in the IceCube
\cite{optim:syn:1474} system.
IceCube relies on a primary site for commitment.
In Telex, each site runs an IceCube engine (or any alternative) to
propose schedules, and the commitment protocol ensures consensus based
on these proposals.
IceCube supports a richer set of constraints and can extract them from
the applications' source code \cite{optim:syn:lan:1476}.

The Ivy peer-to-peer file system \cite{fic:rep:1611} reconciles the
current state of a file from single-writer, append-only logs.
There are several differences between Ivy and Telex.
Ivy is designed for connected operation.
Ivy is state-based and reconciles using a per-byte LWW algorithm by
default.
Whereas Telex localises logs per document, in Ivy there is a single
global log for all the updates of a given participant.
Reading any file requires scanning all the logs in the system, which
does not scale well, although this is offset somewhat by caching.
Ivy has no commitment protocol, therefore a state may remain tentative
indefinitely.

The Ivy authors suggest that malicious updates can be removed after the
fact, by ignoring the corresponding log.
However, since Ivy does not record constraints, it cannot reconstruct a
correct state: for instance, an update by an innocent user that depends
on a previous but malicious update cannot be removed.

\section{Conclusion}
\label{sec:conclusion}

We presented the Telex system for shared mutable documents in a
distributed system.
We presented our motivations, its formal principles, the engineering
design and implementation, and a number of prototypical applications.
We also provided some performance measurements.

Our two main innovations are our principled approach based on
action-constraint graph, and the multilog structure.
The former enables Telex to provide correctness guarantees while
maintaining application concurrency invariants.
It also allows a clear separation between the responsibilities of
applications, and those of the system.
Thanks to constraints, applications specify precisely the level of
consistency that they need, and the system enforces that level
efficiently, and no more.

Independently of the ACG, we argue that the multilog structure is better
adapted to shared, mutable documents than ordinary files, especially in
a collaborative environment.
A file system may provide guarantees for directories, but generally only
best-effort consistency for files.
Furthermore, the design goals of a file system are likely to be
different from the needs of actual applications.

The multilog structure decouples reads and writes, avoids contention,
encourages locality, and allows efficient linear access.
Software at a higher level interprets the logs to reconstruct the
application state.
In our case, this is Telex, but it could be the application directly.
Multilogs do not impose any unnecessarily limitations.

Telex is open source software, available at
{\url{gforge.inria.fr/projects/telex2}}.

\subsection*{Acknowledgments}

We thank Abhishek Gupta, of Indian Institute of Technology Guwahati, for
implementing multilogs during an internship at INRIA and for authoring
the SWA application, and Zenon Perisé, of Universitat Oberta de Catalunya,
for authoring the Collaborative Environment application.

{\scriptsize
\bibliographystyle{plain}
\bibliography{bib,shapiro-bib,local}
}
\end{document}